\documentclass[twocolumn, prc, showpacs, amsmath, amssymb, epsfig, 10pt]{revtex4}
\usepackage{epsf, epsfig, graphicx}
\newcommand{\be}{\begin{equation}}
\newcommand{\ee}{\end{equation}}
\newcommand{\bea}{\begin{eqnarray}}
\newcommand{\eea}{\end{eqnarray}}

\begin{document}

\title{ Sound Wave in Hot Dense Matter Created in Heavy Ion Collision }
\author{ X. Sun$^{a,b}$, Z. Yang$^{c,d}$\\
$^a$Institute of High Energy Physics,  Beijing 100049,  China\\
$^b$Graduate University of the Chinese Academy of Sciences,  Beijing 100049,  China\\
$^c$Department of Engineering Physics,  Tsinghua University,
Beijing 100084,  China\\
$^d$Center of High Energy Physics,  Tsinghua University,  Beijing
100084,  China}
\begin{abstract}
A simple model is proposed to study the sound wave in hot dense
matter created in heavy ion collisions by jet. The preliminary
data of jet shape analysis of PHENIX Collaboration for all
centralities and two directions is well explained in this model.
Then the wavelength of the sound wave, the natural frequency of
the hot dense matter  and the speed of sound wave are estimated
from the fit.
\end{abstract}

 \maketitle

%\noindent ${\bf PACS: 25.75, 23.70.E}$

\section {Introduction}
The energy loss of hard partons induced at the early stage of
heavy ion collision has been regarded as a promising tool to study
the properties of deconfined matter that created in the
untrarelativistic heavy ion
reactions\cite{renk1,Jet1,Jet2,Jet3,Jet4,Jet5,Jet6}. While the
energy and momentum that lost by a hard parton propagating through
hot and dense matter would obviously be redistributed in the
created medium, recent results\cite{PHENIX-2pc} of the
measurements of two-particle correlations, which involve on hard
trigger particle, have shown us a surprising features that double
peaks are measured on the away side for all non-peripheral
collisions for Au+Au collisions. This distinguished feature is
quite different from that for Au+Au peripheral collisions and p-p
or d-Au collisions. This surprising feature attracts interests of
many physicists and a lot of models have been proposed to explain
this phenomenon, e.g., possible colored sound
mode\cite{renk1,ruppert1,ruppert2,ruppert3,ruppert4} proposed by
J. Ruppert and B. M\"uller \emph{et al}(see\cite{ruppert5} for
detailed review by J. Ruppert).

Among these attempts, the idea of conical flow
\cite{Casalderrey-Solana:2004qm}, similar to  sonic booms induced
by supersonic bodies, is regarded as a promising explanation. In
fact, this idea has been proposed before the double peaks are
observed definitely and it predicts this distinguished phenomenon.
 The most excellent property of this model
is introducing a collective exciting mode to the created medium
after heavy ion collision. In conical flow model, in order to
separate the conical flow and the elliptic flow, it is convenient
to focus on the most central collisions, where the elliptic flow
is as small as possible. The formalism used in this model is based
on linearized hydrodynamics,which holds true in the case the total
energy-momentum density deposited by the jets is a small
perturbation compared to the total energy of the medium when
describing medium far enough from the jet head.The assumptions
that the perturbed medium is homogeneous and at rest are also
required. Although comparing with the angular correlation pattern
of hard hadrons  shows good agreement with theory and
experiment,the shortcoming of the conical flow model is obvious.

To study  dependence of the angular correlation pattern to
centrality i.e. the impact parameter,this letter derives the merit
of conical flow model and uses a more microscopic and basic
formalism of this effect. The collective exciting mode means the
perturbation induced by the jet is not separately. There is
correlation between contiguous perturbations. In our model, we
introduce the intrinsic
 sound  wave of  relativistic hot dense matter which will be
discussed in detail in the following to express the correlation
between perturbations.

  The letter is organized as follows: In Section \ref{sec2} we
explain basic framework  and assumptions in our model. In Section
\ref{sec3} this result of the model is discussed and compared with
experiment data. A summary is made in Section \ref{sec4}

\section {framework  and assumptions}\label{sec2}
For the shortage of knowledge of interaction between parton and
gluon, it is difficult to know the behavior of hot and dense
matter created in heavy ion collision. However, to study the
energy losses of fast quenched parton, one treats two different
types of cases separately : (i) the radiative losses, producing
relativistic gluons; and (ii) the scattering/ionization losses,
which deposit energy and momentum directly into the medium. Both
of them have relations to the property of the hot dense matter
created in heavy ion collision.

To overcome this shortage, one can understand this energy
transportation on the point of view of a wave. Generally, a wave
can travel through  whatever matter which satisfies $\partial
P/\partial V <0$ with $P$ and $V$ being pressure and volume of the
matter respectively,like lattice wave in solid, sound wave in
liquid and gas,carrying energy.

 It is reasonable to deem  $\partial P/\partial V <0$ is true for hot dense matter
 produced in heavy ion collision since it is wrong only for undetected  dark matter which is important in
cosmology. If a wave travelling the matter freely, the velocity
and frequency of it  are intrinsic properties of the matter. The
hot dense matter created in heavy ion collision can be regarded as
a kind of liquid or gas, so we call the wave travelling through it
as sound wave. And the perturbation induced by fast parton will
propagate freely until it reaches  the boundary of the source. So
the frequency the sound wave should be the natural frequency of
the hot dense matter.

As a standard wave equation always successes in describing  wave
phenomena in  having equilibrated system, we  assume it  to be
able to describe the sound wave in hot dense matter for a
admittedly thermalization can be reached after heavy ion
collision. The speed of sound $c_s$ is the unique parameter of the
wave equation and one parameter in our model which should be
invariant in every case of centricity. Then sound wave travelling
the hot dense matter  obeys a wave
 equation reading
 \be\label{wave}
\frac{\partial u^2}{\partial t^2}-c_s^2\triangle u=0
 \ee
where $c_s$ is the speed of sound of the medium produced in heavy
ion collision. We note that equation (\ref{wave}) is only a
approximation of true one  without considering anharmonic terms,
which is reasonable where  the anharmonic terms inducing dissipative
effect in the medium is negligible.  This approximation is   also
necessary for Mach-Cone description and its reasonability  should be
tested in further study of non-perturbation QCD. However,equation
(\ref{wave}) at least  holds true where is away from the trajectory
of quenched fast parton for a distance of order of  sound
attenuation length ,$\Gamma = (4/3)\eta/(\epsilon+ p)$, with $\eta$
being the shear viscosity.

On the other hand ,as the quenched fast parton is constantly
emitting gluons, which emit new ones etc., the whole shower is a
complicated nonlinear phenomenon \cite{Casalderrey-Solana:2004qm}.
The multiplicity of this shower grows nonlinearly with time.
Eventually,  the core may become a macroscopic co-moving  body.
All perturbations originating from this co-moving system produced
by the quenched fast parton requires the consideration of  Doppler
effect. The velocity of the co-moving system $c_c$ is another
parameter of our model which should be determined in experiment
fit. However, it can also be discussed qualitatively. $c_c$ should
increase with the increment of the quenched parton and decrease
with the increment of the length of the hot dense matter through
which the quenched parton travels. We note that the source is
expanding instead of static,so the whole medium through which
sound wave travels has a radial or elliptical velocity which
should affect the energy distribution of the final-state
particles. This effect is approximately  negligible in
center-center collision because a radial velocity field  will not
change the spatial angle distribution of energy. For a simple
case, we adopt a radial velocity  field in our discussion.

Now we depict our image of understanding the interference of
perturbation induced by the quenched fast parton in Fig.1,
%%%%%%%%%%%%%%%%%%%%%%%%%%%%%%%%%%%%%%%%%%%%%%%%%%%%%%%%%%%%%%%%%%%%%%%
%%%%%%%%%%%%%%%%%%%%%%%%%%%%%%%%%%%%%%%%%%%%%%%%%%%%%%%%%%%%%%%%%%%%%%%
\begin{figure}[ht]
\vspace*{+0cm} \centerline{\epsfxsize=6cm \epsffile{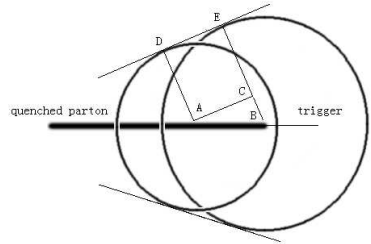}}
\caption{\label{Fig.1} diagrammatic sketch of perturbation
interference induced by quenched fast parton. perturbations
originates from A and B has a phase difference $2\pi
L_{\overline{BC}}/\lambda$ where $\lambda$ is the wavelength of
sound wave traveling through the medium created in heavy ion
collision }
\end{figure}
%%%%%%%%%%%%%%%%%%%%%%%%%%%%%%%%%%%%%%%%%%%%%%%%%%%%%%%%%%%%%%%%%%%%%%%%
where perturbations originate from  line section  $\overline{AB}$
with phase $2\pi l c_s/(V \lambda)$ where $l$ is the distance
between $B$ and point located in line section  $\overline{AB}$, $V$
is the velocity of the quenched fast parton ,always taking $c$ for
it, and $\lambda$ is the wavelength of the sound wave. The
perturbations interfere with  phase and produce the interference
pattern ,for example, in  $D$ and $E$. This pattern represents the
distribution of energy of perturbations and  should be equivalent in
shape to two-particle correlation about jet which is measured in
experiment.

%The shortage of this model is that

 The momentum distribution of
particles induced by the quenched fast parton is not treated in
this model since it has relations to durative scattering rather
than collective excitation mode. For this reason, we can only
adjust the relative intensity between interference pattern of
trigger and its counterpart according to experiment.

Conclusively, there are three parameters in our model totally,
which are the speed of sound $c_s$, the velocity of co-moving
system $c_c$ and the ratio of the  size of the source to the
wavelength of the sound wave $L_{\overline{AB}}/\lambda$. Because
the system in which we are interested evolutes with a finite
lifetime, these parameters should be regarded as time-weighted
averages respectively.

\section {result}\label{sec3}

In experiment, one can measure the  azimuthal correlation ,around a
high momentum trigger, of particles within a special momentum window
, which should be reproduced by the diffraction pattern of the
carrying energy sound wave in our model.

The perturbations induced by a same parton should be coherent by
phase, while these induced by different partons should be superposed
by intensity. Thus the diffraction pattern for one parton reads
 \bea\label{intensity}
I(\theta)=A\left|\frac{\sin[\pi s(\cos\theta-c_s)/\lambda]}{\pi
s(\cos\theta-c_s)/\lambda}\right|^2
 \eea
 where  $\theta$ is the azimuth angle in the frame of center of
 mass and $s$ is the length through which the fast parton passes
 and $\lambda$ is the wavelength of the sound wave.
 For showing the result in laboratory frame, one can substitute
 $\theta'$ for $\theta$ with,
 \bea
  \theta'=\left\{\begin{array}{ll}
                 \tan^{-1}\left(\frac{p\sin\theta}{\gamma(p\cos\theta-\beta E)}\right),
       & p\cos\theta>\beta E, \\
                 \tan^{-1}\left(\frac{p\sin\theta}{\gamma(p\cos\theta-\beta E)}\right)+Sign(\theta)\pi,
       & \texttt{otherwise}.
               \end{array}
         \right.
  \label{translate}
 \eea

where $\beta$ is $c_c$ and $\gamma=(1-\beta^2)^{(-\frac{1}{2})}$.
$p$ and $E$ in (\ref{translate}) means the momentum and energy of
observed  particle in laboratory frame. The analytic form of
$I(\theta')$ is  difficult to write out, though it looks very easy.

The speed of sound $c_s$ should be a constant for a fixed
temperature and the speed of co-moving system  $c_c$ should increase
obviously with the momentum of trigger particle. the length of
trigger passing  $s$ should be determined by the centricity of the
collision, i.e. the multiplicity of the collision. Fig.1 shows the
comparison between preliminary data of jet shape analysis of PHENIX
Collaboration\cite{Akiba:2005bs} and our calculation. Different
marks denote experiment results in different centralities, while
different styles of lines denote our calculations using different
parameters. The momentum of particle used in the calculation, 2.5
GeV , is the average of the momentum range of experimentally
measured particles i.e. 2-3 Gev. The used speed of sound  ,
 much smaller than that used in Mach cone calculation in
 \cite{Casalderrey-Solana:2004qm} where $c_s=0.33c$, is $0.129c$ for the better
 fit, which is a constant in calculation.
In this case the peak  of induced particle should be at angles about
1.7 radian relative to the trigger. The normalization for all
centralities has been performed.
\begin{figure}[ht]
\vspace*{+0cm} \centerline{\epsfxsize=6cm \epsffile{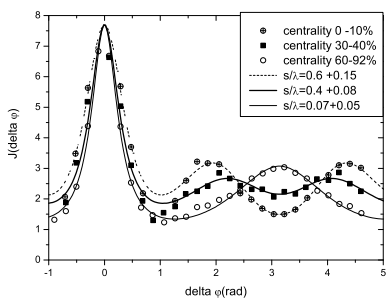}}
\caption{\label{Fig.2} the comparison between experiment and theory,
The dash line corresponds to center-center collisions, where the
size of source is relative large. The thin corresponds to peripheral
collisions. The particle momentum is 2.5 Gev and $c_s=0.129$}
\end{figure}

The parameters in the calculation is listed below

\bea
\begin{tabular}{|c|c|c|c|c|c|c|}
  \hline
   & 0-10\%$\uparrow$& 60-92\%$\uparrow$& 30-40\%$\uparrow$&0-10\%$\downarrow$& 30-40\%$\downarrow$& 60-92\%$\downarrow$ \\
  \hline
  $s/\lambda$ & 0.15& 0.08&0.05&  0.6& 0.4& 0.07 \\
  \hline
  $c_c$ & 0.92  & 0.95  & 0.96& 0.2 & 0.3& 0.59 \\
  \hline
\end{tabular}\nonumber
\eea.

The $\uparrow$ means the forward jet i.e. the jet caused by the
trigger particle while $\downarrow$ means the backward jet which is
caused by the  counterpart of the trigger.

It is easy to understand the variation of the parameters. The
centrality of the collision , relative to the source size, is
reflected by the $s/\lambda$. As the increment of the source size,
the co-moving velocity decreases.This reflects the energy loss of
particle when traveling the source.

The intensity ratios between forward jet and  backward jet are 0.60,
0.51 and 0.67 for $0-10\%,30-40\%,60-92\%$ respectively, which come
from data fitting and will not be discussed in our model as is above
mentioned.

Now the picture of sound wave in hot dense matter is to some extend
obvious. Let's take double HBT radius\cite{Adams:2004yc} , $5 fm$,
as the source size for center-center collision, the wavelength of
sound wave in hot dense matter created in heavy ion  collision is
estimated as $\lambda=(2\times 5 fm)/(0.6+0.15)=13.3 fm$. The
natural frequency of the hot dense matter is estimated as
$c_s/\lambda=2.91\times 10^{21}Hz$. These quantities will be
constants at fixed temperature.

But Making conclusion that we have catch signals about the intrinsic
eigenvibration of hot dense matter is premature at present. After
all, there must be other models being able to explain the same data.
To test the existence of the sound wave, we can examine other
behaviors of this model.

The velocity of co-moving system $c_c$ will increase as the
increment of the  momentum of the trigger. For higher momentum of
trigger particle,the distance between two peaks in backward jet in
center-center collisions will decrease since the Lorentz
transformation in (\ref{translate}), which can be tested in LHC
where a lot of jets with high momentum trigger will be produced.

We can also make a more fine bins of centrality to draw the jet
induced particle azimuth correlation again, which will give a more
crashing test of this model.

Essentially, the properties of the sound wave are determined by the
interaction between particles in hot dense matter. One can conceive
that it is quite different for confined and deconfined  scenarios.
So it is meaningful to compare different scenarios to experiment,
which will be a interesting tool to research QGP.

\section {conclusion}\label{sec4}
We try to study the sound wave in hot dense matter created in
heavy ion collisions in this letter by jet. The distribution of
final-state particle is described by the interference of the sound
wave induce by the quenched fast parton. The productions of a
co-moving system  around the quenched fast parton is introduced.
The preliminary data of jet shape analysis of PHENIX Collaboration
for all centralities and two directions is well explained in our
model . From the parameter of the result, we estimate the
wavelength of the sound wave is 13.3 fm, the natural frequency of
the hot dense matter is $2.91\times 10^{21}Hz$ and the speed of
sound wave is 0.129c.

We predict that the distance between two peaks in backward jet in
center-center collisions will decrease since the Lorentz
transformation  for higher momentum of trigger particle, which can
be tested in LHC where a lot of jets with high momentum trigger
will be produced in the future. Our model cares only collective
property of the hot dense matter. Whether the sound wave can
become a tool to test the creation of quark-gluon plasma depends
on the difference between calculations of the sound wave in
quark-gluon plasma scenario and hadron-gas scenario.

\acknowledgements{We thank rof. J. Li and Z. Zhang for the
fruitful discussions. This work was supported in part by the
grants NSFC10447123 and China Postdoctoral Science Foundation of
No. 2004036221.}

\end{document}